\documentclass[sigconf]{acmart}

\def\BibTeX{{\rm B\kern-.05em{\sc i\kern-.025em b}\kern-.08emT\kern-.1667em\lower.7ex\hbox{E}\kern-.125emX}}
    
\usepackage{tabularx,ragged2e}
\usepackage{multirow}
\usepackage{subcaption}
\usepackage{balance}

\setcopyright{acmcopyright}
\copyrightyear{2019} 
\acmYear{2019} 
\acmConference[MM '19]{Proceedings of the 27th ACM International Conference on Multimedia}{October 21--25, 2019}{Nice, France}
\acmBooktitle{Proceedings of the 27th ACM International Conference on Multimedia (MM '19), October 21--25, 2019, Nice, France}
\acmPrice{15.00}
\acmDOI{10.1145/3343031.3356084}
\acmISBN{978-1-4503-6889-6/19/10}

\begin{document}
\fancyhead{}

%
\title{SMP Challenge: An Overview of Social Media Prediction Challenge 2019}

\author{Bo Wu}
\email{bo.wu@columbia.edu}
\affiliation{%
  \institution{Columbia University}
}

\author{Wen-Huang Cheng}
\email{whcheng@nctu.edu.tw}
\affiliation{%
  \institution{National Chiao Tung University}
}

\author{Peiye Liu}
\email{liupeiye@bupt.edu.cn}
\affiliation{%
  \institution{Beijing University of Posts and Telecommunications}
}

\author{Bei Liu}
\email{bei.liu@microsoft.com}
\affiliation{%
  \institution{Microsoft Research Asia}
}

\author{Zhaoyang Zeng}
\email{zengzhy5@mail2.sysu.edu.cn}
\affiliation{%
  \institution{Sun Yat-sen University}
}

\author{Jiebo Luo}
\email{jluo@cs.rochester.edu}
\affiliation{%
  \institution{University of Rochester}
}

%

%


%
\begin{abstract}
``SMP Challenge'' aims to discover novel prediction tasks for numerous data on social multimedia and seek excellent research teams. Making predictions via social multimedia data (e.g. photos, videos or news) is not only helps us to make better strategic decisions for the future, but also explores advanced predictive learning and analytic methods on various problems and scenarios, such as multimedia recommendation, advertising system, fashion analysis etc. 

In the SMP Challenge at ACM Multimedia 2019, we introduce a novel prediction task Temporal Popularity Prediction, which focuses on predicting future interaction or attractiveness (in terms of clicks, views or likes etc.) of new online posts in social media feeds before uploading. We also collected and released a large-scale SMPD benchmark with over 480K posts from 69K users. In this paper, we define the challenge problem, give an overview of the dataset, present statistics of rich information for data and annotation and design the accuracy and correlation evaluation metrics for temporal popularity prediction to the challenge. 
\end{abstract}


%

%

%
\maketitle

\section{Introduction}
\begin{figure}[t]
	\begin{center} 
		\includegraphics[width=1.0\linewidth]{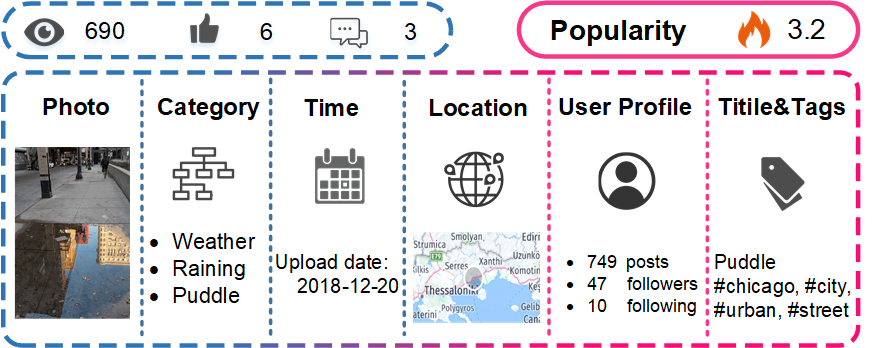} 
	\end{center}   
	\caption{SMP Challenge introduces Temporal Popularity Prediction task for social multimedia. SMPD includes visual content (diverse images with categories into a semantic taxonomy), textual content (e.g. title and custom tags) and spatial-temporal content (e.g. location and time). The popularity score in the figure is prediction target and calculated by ``user interactions'' of online post.}
	\label{fig:post}
	\vspace{-0.22in}
\end{figure}
People are interested in predicting the future. For example, which who will win the upcoming Grammy Awards or which film will be prevalent in next few weeks? Making predictions about the future brings real values to a variety of applications and scenarios~\cite{Martin2016}, such as multimedia recommendation~\cite{Mei2011,Khosla2014,Qian2014}, advertising system~\cite{Li2015,Zhao2015seismic}, fashion analysis~\cite{fashionpop,Hidayati2014}, topic mining~\cite{Ferrara2014,Wu2014learning,Szabo2010predicting} etc.


Therefore, the purpose of SMP Challenge is to discover novel challenge tasks based on numerous resources on social multimedia and seek excellent research teams who are capable of making the prediction. For prediction, the increasing ubiquity of social media (e.g. Facebook, Twitter, Flickr, YouTube, etc.) provides a crucial way for learning about the real-world. Meanwhile, social multimedia data increased interest for researches in study of exploring rich social facts and knowledge with multi-modal information (e.g. images, text, video, events, etc.), while social media is now globally ubiquitous and prevalent. So far the researches of social media prediction covered in several significant areas of multimedia and artificial intelligence, and closely integrated with computer vision, machine learning, natural language and human-centered interaction.

During this year, the task of SMP Challenge 2019 is Temporal Popularity Prediction~\cite{Wu2017DTCN, Szabo2010predicting}, addressing the problem to predict the future popularity of giving posts before they were published in social media. This treats popularity prediction at a time-related prediction problem~\cite{Shulman2016,Kobayashi2016}, and formulates popularity by online attention based on various user interactions (e.g. clicks, visits, reviews). To achieve this goal, the participated teams need to design new algorithms of understanding and learning techniques, and automatically predict with considering post content, future post time and its multiple multimedia information (as shown in Figure~\ref{fig:post}) in a time-related dynamic system~\cite{Myers2014bursty,Kong2014predicting,Yang2011}.

In the literature, several large-scale datasets from social media have been established for various research tasks and helped lead to great advancements in multimedia technology and applications, such as YFCC~\cite{Kalkowski2015}, Yelp2016~\cite{Asghar2016}, Visual Genome~\cite{Krishna2016}, etc. However, most of the existing datasets are limited in the diversity of coverage, i.e. the collected data are often biased to the particular task in question, and lacking cross-task generalization. Therefore, we introduced Social Media Prediction Dataset (SMPD), a large-scale benchmark dataset for sociological understanding and predictions with over 486k posts and 80K users in total. SMPD collects multi-faceted information of a post, such as user profile, photo metadata, and visual content. Particularly, we aim to record the temporal order of social media data. For example, social media posts in the dataset are obtained with temporal information to preserve the continuity of post sequences. Our goal is to make the SMPD as varied and rich as possible to thoroughly represent the social media ``world''.














\begin{figure}[t] 
	\begin{center}
 		\includegraphics[width=0.8\linewidth]{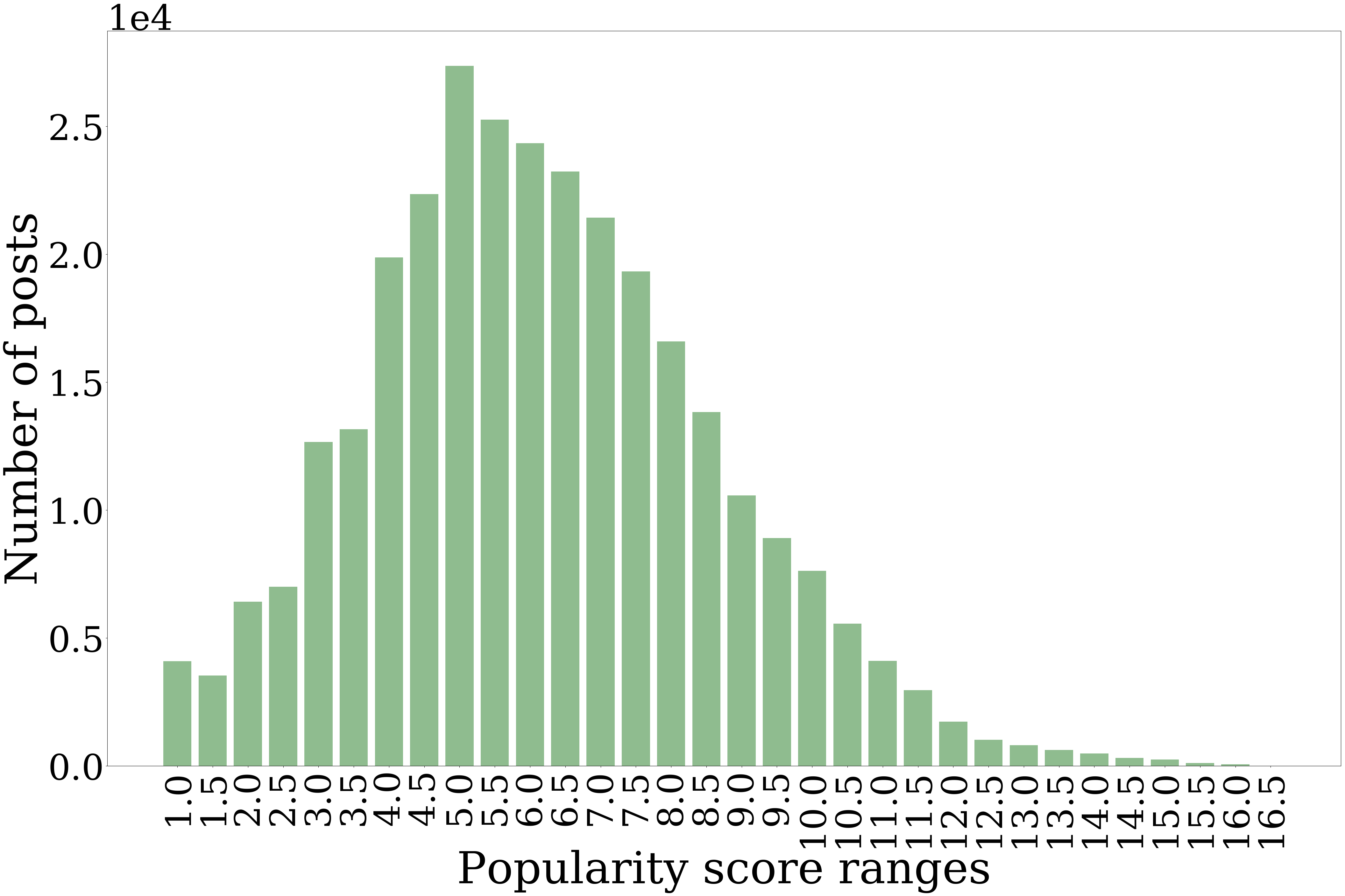} 
	\end{center}
	\caption{The histogram of popularity score. Each score range collects the posts which have a popularity score within the range [x-0.5,x]. }
	\label{fig:his}
	\vspace{-0.22in}
\end{figure}
\section{Temporal Popularity Prediction}
\subsection{Problem Formulation}\label{subsec:formulation}
Temporal Popularity Prediction (TPP) is a novel problem for social media analyzing and learning~\cite{Wu2016TimeMatters}. With the temporal dynamics of the social multimedia system, the popularity of online posts usually changed over time. Influenced by the temporal characteristics~\cite{McGrath1992,Myers2014bursty} with complex contexts or patterns, how to predict accurate temporal popularity become more challenging than before. The task of TPP is to estimate the future impacts of giving social media posts (photos, videos or news) at a specific time before they were shared on the online platform.    
Specifically, given a new post $v$ of a user $u$, predict popularity $s$ describes how many attentions would obtain if it was published at time $t$ on social media. The formulations of popularity can be defined as a score by different dynamic indicators (e.g. views, likes or clicks, etc.) via diverse social multimedia platforms. In our challenge, we use ``viewing count'' as a basic indicator of how popular a post is, while this is more general. So temporal popularity can be defined as the following:

\textbf{Popularity Normalization.} To suppress the large variations among different photos (e.g. view count of different photos vary from zero to millions), we implement a log function~\cite{Wu2016TemporalPrediction} to normalize the value of popularity, based on the previous work, as shown in Figure~\ref{fig:his}. In brief, the log-normalization function for popularity can be defined as:
\begin{equation}
s = log_{2}\frac{r}{d} + 1
\end{equation}
where $s$ is the normalized value, $r$ is the view count of a photo, and $d$ is the number of days since the photo was posted.

Particularly, the post sequence with time information for each of user can be treated as time-series data. SMP Challenge 2019 aimed to make time-series feeds for popularity prediction. Then, we defined sequence data with time orders: 

\textbf{User-Post Sequence.} Suppose we have $n$ user-photo pairs and the sharing time of each pair. Then the user-post sequence can be denoted by $S = \{(u_1, v_1),(u_2, v_2), ..., $ $(u_n, v_n)\}$ with its sharing time order $t_1 \leq t_2 \leq ... \leq t_n$.

\begin{figure}
  \centering\includegraphics[width=3.5in]{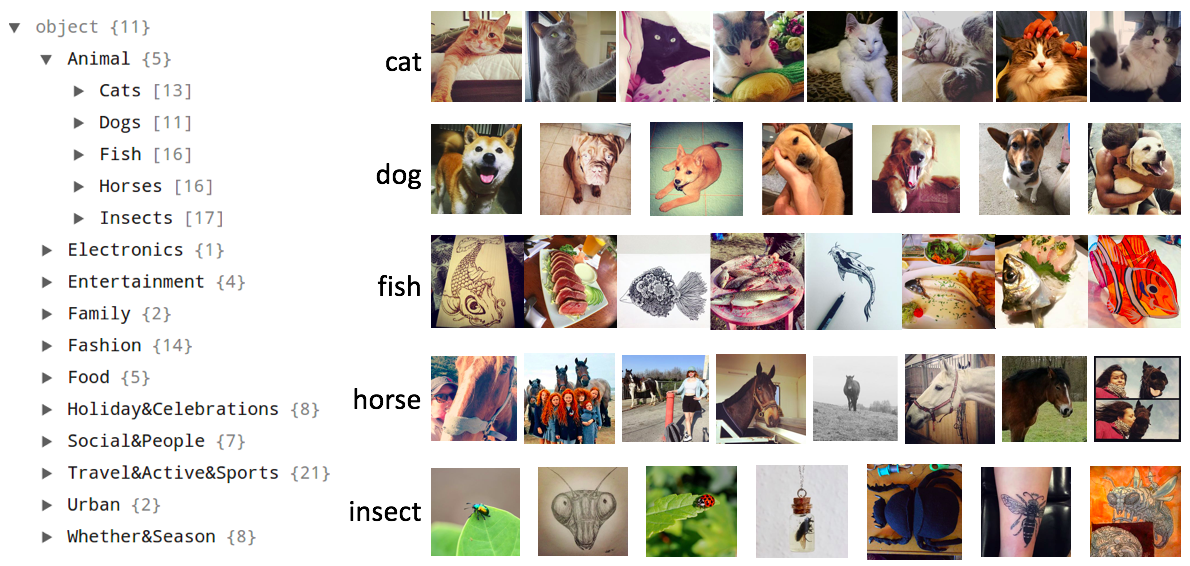}
  \caption{An example of multi-level hierarchical photo categories. It shows a $2^{nd}$-level category ``animal'' with 5 different $3^{rd}$-level categories.}
  \label{examplePic}  
  	\vspace{-0.22in}
\end{figure}

\section{Social Media Prediction Dataset Overview}

Social Media Prediction Dataset (SMPD)~\footnote{http://smp-challenge.com/dataset} is a large-scale benchmark for social multimedia researches. We selected Flickr as the data source of SMPD for multimedia and multi-modal data, which is one of the largest photo-sharing websites with over 2 billion photos monthly\cite{Franck19}. Different with single-task datasets, SMPD is a multi-faced data collection, which contains rich contextual information and annotations for multiple-tasks (such as user profile, post category, customize tag, geography information, photo image, and photo metadata). The overview statistics of the dataset are shown in Table~\ref{tab:sta}. It contains over 486K posts from 69K online users. And each of social media post has corresponding visual content and textual content information (e.g. posted photos, photo categories, custom tags, temporal and geography information).

\textbf{SMPD Building.} 
To create a multi-faced dataset for social media research, we attempt to utilize a concept-based sampling method to collect post data from the search engine of the Flickr platform. The concept-based approach aims to take a tag or concept as a searching keyword, collects the posts that involved with the keyword. On this basis, a second selection will be manipulated to ensure the accuracy of concept-related posts~\cite{Huiskes2008}. 
The advantage of this approach is offering an accurate data source for theme extraction and feature extraction. Unlike traditional social bookmarking, Flickr does not involve creating an explicit vocabulary of tags to describe the post. Therefore, the referencing queries of different categories are filtered from the most popular tags which user liked most in 2015, such as other tag prediction study~\cite{Jang2015}. We filtered the tags within incomplete or typo keywords, such as insta'' or ``instadog'', etc. Then we leave 756 categories within 11 topics for our dataset creation, as shown in Figure 4 and 5. To keep time-orders in our data, we obtained the public post stream continuously for each of the categories in every day from Nov. 2015 to March. 2016. To have various properties in our dataset, we extracted abundant data including visual content (e.g. photo and photo categories), textual content (e.g. post title and custom tags) and Spatio-temporal content, revealing the influence of region and time-zone on online social behavior. Finally, we have a large-scale multi-faced collection. 

\begin{table}
\centering
\caption{Summary of SMPD Statistics.} \label{tab:sta}
\begin{tabular}{l|c|c}%
\hline  %
\multirow{2}*{\textbf{Statistics}}&\multicolumn{2}{c}{ \textbf{Value}}\\
\cline{2-3}
&Train&Test\\
\hline  %
Number of Posts& $3.05\times10^5$&$1.81\times10^5$ \\
Mean popularity of posts& 6.41&5.12\\
STD popularity of posts& 2.47&2.41\\
Number of users &$3.8\times10^4$&$3.1\times10^4$\\
\hline

Number of custom tags& \multicolumn{2}{c}{$2.5\times10^5$}\\
Number of $1^{st}$ level categories& \multicolumn{2}{c}{11}\\
Number of $2^{nd}$ level categories& \multicolumn{2}{c}{77}\\
Number of $3^{rd}$ level categories&\multicolumn{2}{c} {668}\\
Temporal range of posts& \multicolumn{2}{c}{480 days}\\
Average length of title& \multicolumn{2}{c}{29 words}\\
\hline %
\end{tabular}
	\vspace{-0.22in}
\end{table}

\begin{figure}[h]
	\begin{center}
		\includegraphics[width=1.0\linewidth]{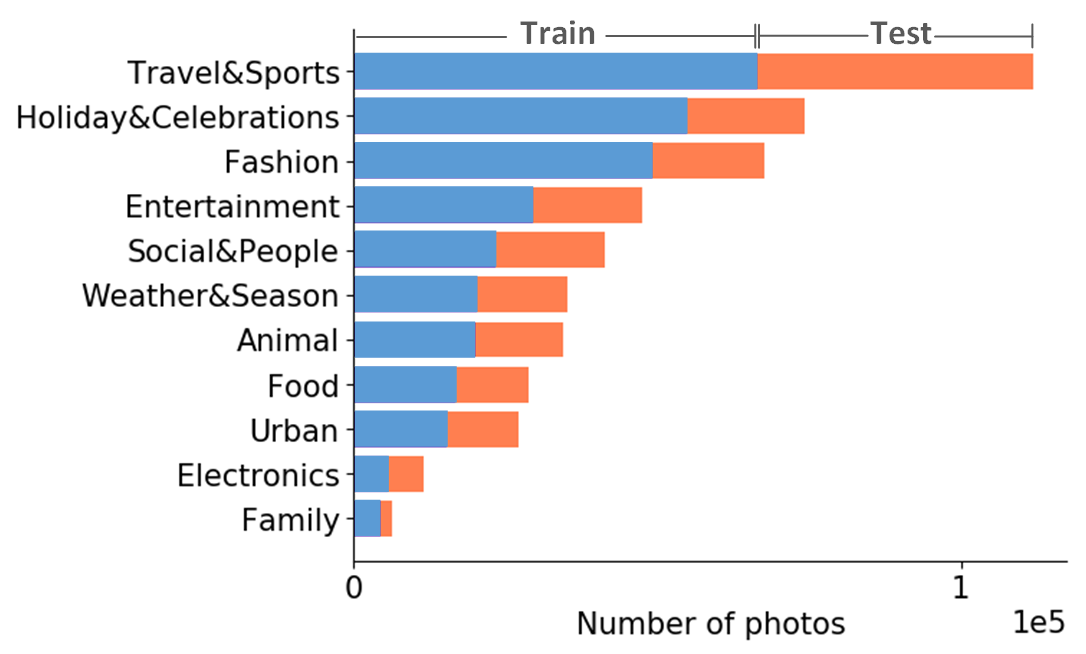} 
	\end{center}
	\caption{The statistics of posts in each $1^{st}$ level category. It shows 11 $1^{st}$ level categories and the number of posts in train and test data.}
	\label{fig:label}
	\vspace{-0.22in}
\end{figure}
\subsection{Visual Content}
 As an old saying: ``A picture is worth a thousand words", it is easier for users to reflect their thoughts or emotions by photo/image on social media~\cite{Cappallo2015}. In our dataset, we collect 486k posts by querying 756 selected key-words (as mentioned in the prior section) with social media APIs~\cite{FlickrAPI}. These key-words can be organized into 11 topics range from nature, people to animal (the directory tree in the left of Figure~\ref{examplePic}). Furthermore, each of key-words represents an individual concept for photo content, such as ``bird", ``flower", etc. In the right part of Figure~\ref{examplePic}, an example category ``Animal" with five different sub-concepts for content visualization is shown. The visual content (photo or image) of the posts with the same key-word are similar in visual view. By utilized these categorized photos, it helps prepare train/test data for computer vision works. Furthermore, we generate 668 individual $3^{rd}$ level human-craft categories for photos of selected posts. By the 3 levels hierarchical fine-grained classifying, SMPD provides fine-grained classes.




\subsection{Spatio-temporal Content}

\subsubsection{Time}
\begin{figure}[t]
	\begin{center}
		\includegraphics[width=1.0\linewidth]{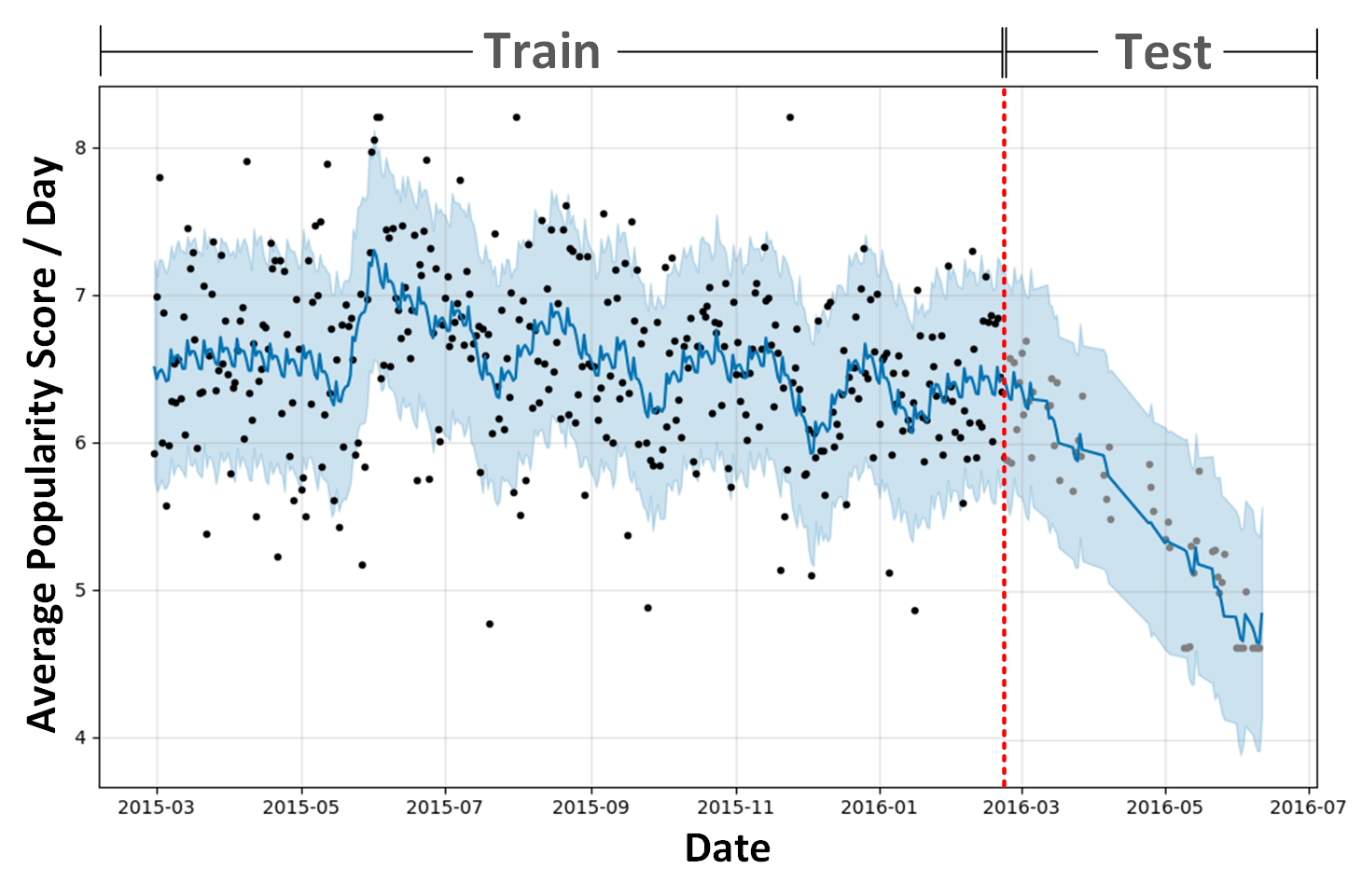} 
	\end{center}
	\caption{Overall distribution of popularity over days. The discrete black points represent the average popularity score per day. The blue line and shadow interval represent a regression result about the score. Specifically, the left part of the red line represents the training data and the right part represents the test data.}
	\label{fig:average}
	\vspace{-0.22in}
\end{figure}

Popularity prediction of social media posts is a time-sensitive task~\cite{Ellering2016}. Temporal context of posts records user activities, and it is necessary to identify the uploading time. Meanwhile, Flickr provides an uploading time for each submitted post. Figure~\ref{fig:average} plots the average post counts and its uploaded months in our dataset. 
The posts show the most of posts in SMPD were uploaded between March 2015 to the creation time of the dataset in 2016. Among the upload dates, October and December become the popular month for sharing posts on the Flickr social media platforms. We attribute those improvements to the holidays and the end of the year.

\subsubsection{Location}
\begin{figure}[h]
  \centering\includegraphics[width=1.0\linewidth]{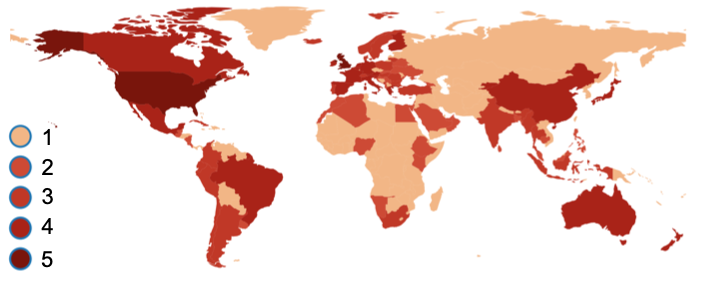}
  \caption{The distribution of posts around the world. The legend in the figure represents 5 equal interval ranges of Log-normalization value of posts$'$ number in each territory from 0 to 9.28. The deeper the color is, the more posts are posted in the corresponding territory.}
  \label{fig:maps}  
  	\vspace{-0.22in}
\end{figure}

Location information provides user spatial distribution~\cite{Yang2016}. Not all posts have location information, but the location information of photos point out the spatial region of user activities. In SMPD, 32,068 posts have POI (Point of Interest) location information with a geographic coordinate, either manually by the user or automatically via GPS. Using this information, we were able to map 10\% of all items in the dataset to a single country or area. Furthermore, SMPD provides geo accuracy value to represent the accuracy level of the location information, which ranges from 1 to 16 and represents from words level to street-level accuracy. To suppress the large variations among the number of posts in different territories, we implement a log function to normalize the number of posts in each country, based on the previous work.
The distribution of all items over territories is shown in Figure~\ref{fig:maps}.

\subsection{Textual Content}
In addition to visual content, we also collected the surrounding text of posts provides to show semantic information for each post. As statistics, there are more than 95\% posts have relative descriptions or titles. When uploading a photo on the social media platform, the relative textual content is appended to provide more details about the photo content or publisher status. 

\textbf{Post Title.} Each posted photo has a unique title named by the user. As the saying goes: ``There are a thousand Hamlets in a thousand people$'$s eyes", each title contains the explanation and understanding of the photo. As shown in Table~\ref{tab:sta}, users utilize average 29 words to describe the content of uploaded photos, which not only helps to analyze the visual content of the posts but also relate to the popularity of the corresponding post.
\begin{figure}[ht]
	\begin{center}
		\includegraphics[width=1.0\linewidth]{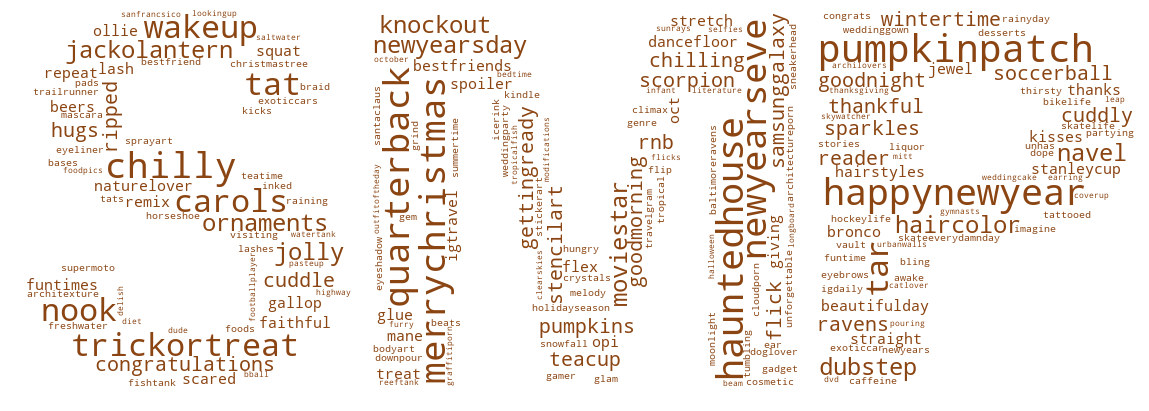} 
	\end{center}
	\caption{The tag-cloud of 668 $3^{rd}$-level keywords of photos. The larger the font size used, the corresponding tag is more frequently used.}
	\label{fig:cloud}
	\vspace{-0.22in}
\end{figure}

\textbf{Post Tags.} Most of social network sites provide hash-tags to make user easier to find relevant post by topic with the same tag. It is possible to label a single post with multiple tags. With this kind of flexibility, this method is easier than the traditional one-to-one classification. By counting the tag frequently used, we can have a glimpse of which topics are more popular within these social network sites. In Figure~\ref{fig:cloud}, we generated the ``tag cloud" of post tags. The larger the font size used, the corresponding tag is more frequently used. Based on the figure, the users of Flickr prefer to share a holiday-related popular tag. 

\section{Evaluation}
To measure the performance of temporal popularity prediction on time-series data, we adopt the time-related partition strategy to generate train/test sets for evaluation. In proposed time-series data SMPD, we have a 10 length of the time window to build user-post sequences and divides the sequences to 2:1 for training and test dataset. Specifically, the train and test sets also share similar numbers of users.

By objective evaluation, we measure the performance of submitted methods on the unpublished SMPD test set. Our evaluation protocol is applied to the following criteria:
\begin{itemize}
\item Ranking Relevance: to measure the ordinal association between ranked predicted popularity scores and actual ones.
\item Prediction Error: to judge the error of the score prediction.
\end{itemize}
As quantitative metrics of performance evaluation, we will compute Spearman Ranking Correlation (SRC, or Spearman$'$s Rho), Mean Absolute Error (MAE) for each submitted model. SRC is a nonparametric measure of rank correlation, it applied to measure the ranking correlation between ground-truth popularity set $P$ and predicted popularity set $\hat P$, varying from 0 to 1. If there are $k$ samples, the SRC can be expressed as:
\begin{equation}
SRC= \frac{1}{k-1} \sum ^k _{i=1} \left( \frac{P_i - \bar{P}}{\sigma_P} \right) \left( \frac{{\hat{P}}_i - \bar{\hat{P}}}{\sigma_{\hat{P}}} \right), 
 \end{equation}
where $\bar{P}$ and $\sigma_P$ are mean and variance of the corresponding popularity set. Furthermore, we also use Mean Absolute Error (MAE) to validate the prediction error. The goal of MAE is to calculate the averaged prediction error: 
\begin{equation}
MAE=\frac{1}{k}\sum_{i=1}^{n}\mid\hat{P}_{i}-P_{i}\mid.
 \end{equation}
 
The ranking for the competition is based on an objective evaluation. Specifically, a rank list of teams is produced by sorting their performance on each of objective evaluation metrics, respectively. The final rank of a team is calculated by combining its two ranked metrics for balance. The smaller the final ranking, the better the performance.

\section{Conclusions}

In this paper, we have presented an overview of SMP Challenge 2019 and proposed a large-scale social multimedia dataset for real-world prediction challenges. Meanwhile, we formulate the temporal popularity prediction task, analyzes the proposed dataset and define evaluation metrics. You can find more information about the task, dataset and challenge at SMP Challenge website~\footnote{http://www.smp-challenge.com}.

%
\begin{acks}
We would like to thank CAS-ICT, Microsoft Research Asia, Academia Sinica for their helpful support. SMP Challenge 2019 was supported by Columbia University. SMP Challenge 2017 and 2018 was sponsored by Kwai Inc. and JD AI Research. 
\end{acks}

%

\bibliographystyle{ACM-Reference-Format}
\balance
\bibliography{smpchallenge} 

%

\end{document}